\documentclass[fleqn,usenatbib]{mnras}

\usepackage{newtxtext,newtxmath}
\usepackage{hyperref}

\usepackage[T1]{fontenc}

\DeclareRobustCommand{\VAN}[3]{#2}
\let\VANthebibliography\thebibliography
\def\thebibliography{\DeclareRobustCommand{\VAN}[3]{##3}\VANthebibliography}

\usepackage{graphicx}	
\usepackage{amsmath}	

\title[Physics of Cometary Anti-tails]{The Physics of Cometary Anti-tails as Observed in 3I/ATLAS}

\author[Keto \& Loeb]{
Eric Keto,$^{1}$\thanks{E-mail: eketo@cfa.harvard.edu (EK)}
Abraham Loeb,$^{1}$
\\
$^{1}$Department of Astronomy, Harvard University, 60 Garden Street, Cambridge MA 02138\\
}

\date{Accepted XXX. Received YYY; in original form ZZZ}

\pubyear{2025}

\begin{document}
\label{firstpage}
\pagerange{\pageref{firstpage}--\pageref{lastpage}}
\maketitle

\begin{abstract}
Observations of interstellar comet 3I/ATLAS at 3.8 au show an elongated coma similar to a cometary tail but pointing in the direction of the Sun, possibly the first instance of this type of anti-tail which is not a result of perspective.  We explain the anti-tail as an anisotropic extension of the snow line, or survival radius of a sublimating ice grain, in the direction of the Sun. The anisotropy is due to the difference in the sublimation mass flux in the solar and perpendicular directions caused by the change in the illumination angle of the cometary surface.  The stronger sublimation mass flux in the solar direction results in ice grains with larger sizes, longer sublimation lifetimes, and a snow line at a larger radial distance with respect to other directions. The observed radial surface brightness profiles as a function of illumination angle are well reproduced by a Haser-type radial outflow with constant velocity and sublimating ice grains with angularly dependent survival lengths.
\end{abstract}




\section{Introduction}

Observations of the interstellar object, 3I/ATLAS, at a distance of 3.8 au from the Sun, by the Hubble Space Telescope (HST) on July 21, 2025 \citep{Jewitt_2025} revealed an anti-tail in the solar direction that is not due to projection. 
Confirmed by other observations \citep{Tonry_2025,Bolin_2025,Cordiner_2025,Chandler_2025,Seligman_2025},
this phenomenon is not common and possibly observed for the first time.  
 In this paper, we explain the development of an anti-tail based on simple physical models of comets. 
 
 We model the coma as a radial outflow of sublimating ice grains ejected by gas drag off the
 cometary surface.  Calculated surface temperatures and grain survival times indicate that the sublimation 
 mass flux off the cometary surface must be predominantly CO$_2$ as observed
 \citep{Lisse_2025,Cordiner_2025}, while the grains must be H$_2$O ice. The radial extent of the coma
 depends on the distance that an ice grain survives while sublimating. This snow-line length
 varies with the solar illumination angle principally through the sublimation mass flux which determines
 the maximum liftable grain size \citep{Whipple_1951} and the maximum of the grain size distribution. 
From a chain of simple calculations, we determine the maximum ice-grain size and terminal velocity as a function of the illumination angle to derive the survival length, $\ell = v_\infty t_{\rm life}$ as a function of illumination angle.

 \section{The anti-tail in the HST image}
 
\begin{figure}
\includegraphics[width=3.0in,trim={0 0.0in 0.0in 0.75in},clip]{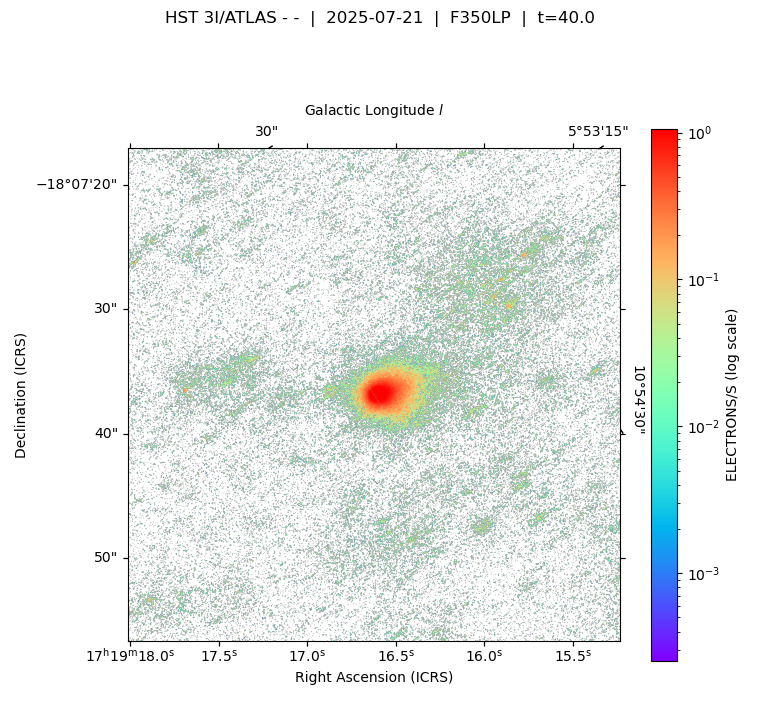} 
\caption{
Left: HST image in log$_{10}$ counts (electrons s$^{-1}$). 
}
\label{HST_image}
\end{figure}

\begin{figure}
\includegraphics[width=3.5in,trim={0 0.0in 0.0in 0.0in},clip]{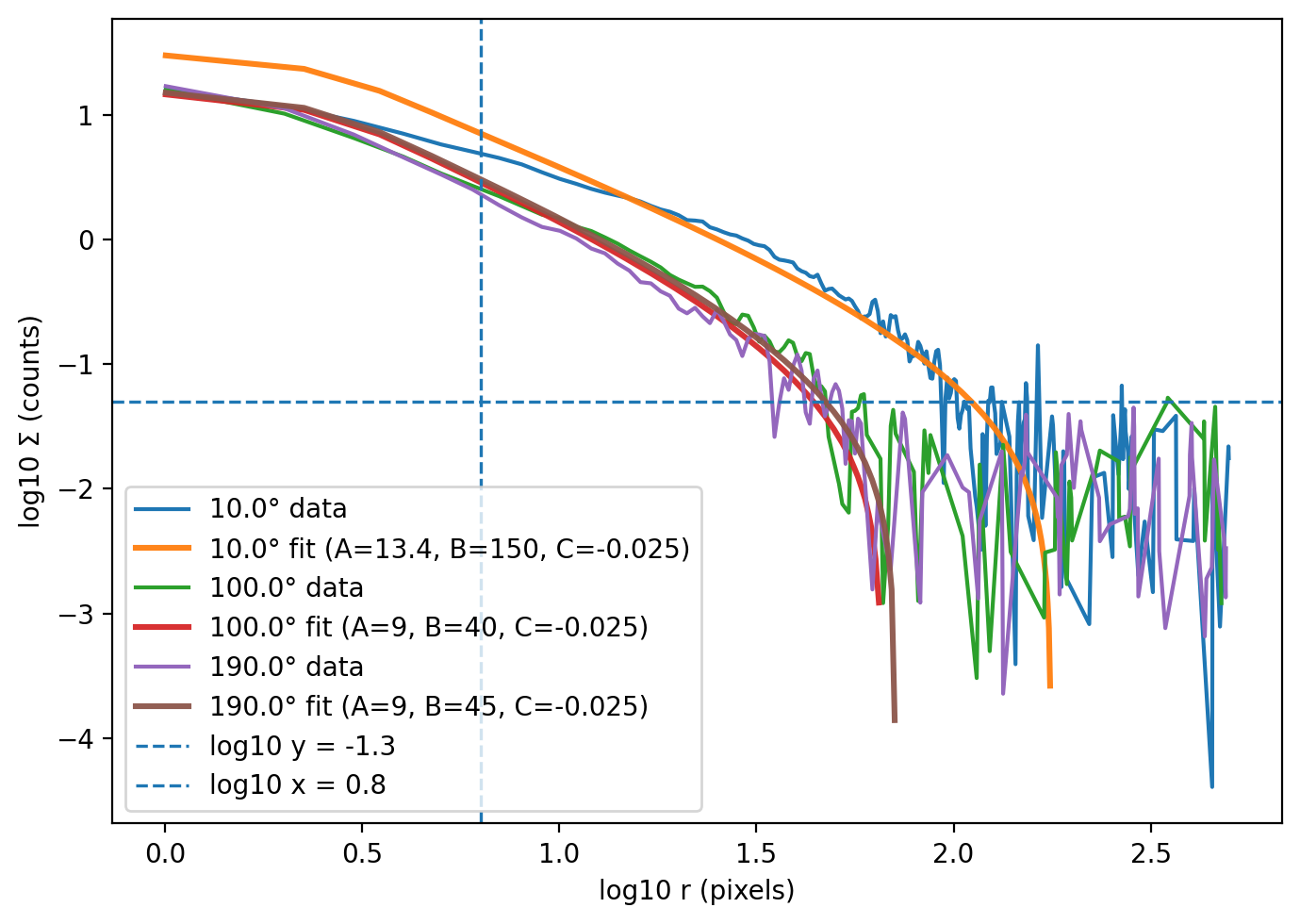} 
\caption{
Radial profiles of the surface brightness at angles of 10$^\circ$, 100$^\circ$, and 190$^\circ$
where the angles are measured with respect to the $x,y$ coordinate axes. The observed profiles are
overlaid with fits to a model Haser-type surface brightness. The $x,y$ axes have units of pixels (0.04") 
and counts (electrons s$^{-1}$).
}
\label{Haser_profiles}
\end{figure}

Figures \ref{HST_image} and \ref{Haser_profiles} show two representations of the HST data. The image is shown with a logarithmic color
transfer function. Radial profiles are shown along directions of 10, 100, 190$^\circ$ with respect to the coordinate axes.
The profile of 10$^\circ$, in the direction of the Sun, is longer than the profiles in the two perpendicular directions
at 100 and 190$^\circ$. 
None of the profiles fits a single-slope power law as would be expected from a radial dilution of the number density
in a constant-velocity, spherical outflow. 
Their curvature suggests an exponential decrease in the number density with distance from the nucleus
as previously mentioned \citep{Jewitt_2025}.

We can model the curvature of the radial surface brightness profiles with a Haser-type density law \citep{Haser_2020} that develops from 
flux conservation in a constant-velocity, spherical outflow combined with destruction of the 
particles,
\begin{equation}\label{haser1}
\frac{1}{r^2} \frac{d}{dr} [ r^2 v(a) n(r,a)]  = -\frac{n(r,a)}{t_{\rm life} } . 
\end{equation}
where $a$ is the ice grain size with velocity, $v(a)$ and  lifetime $t_{\rm life}$.
The solution for the number density modifies the solution for a steady-state outflow by an exponential, 
\begin{equation}
n(r,a) \propto r^{-2} \exp\bigg( -\frac{r} {\ell(a)}\bigg) .
\end{equation}
If we select a representative ice grain size to replace the ice-grain size distribution,
we can use the same integral solution for the surface brightness as in the original Haser solution for 
molecular emission.
 The surface brightness follows,
\begin{equation}\label{haser3}
\Sigma(\rho,a) \propto \frac{Q} {4\pi v(a) \rho} 2 K_0( x )
\end{equation}
where $\rho$ is the projected radius, $Q$ is the mass flux of the outflow and 
$K_0$ is the modified Bessel function of the second kind, and $x=\rho / \ell (a)$.

Figure \ref{Haser_profiles} shows the mathematical fits to the Haser  model,
\begin{equation}
I(x) = C + (A/x) \exp{ (x/B) }
\end{equation}
where $I(x)$ is the observed counts at coordinate $x$ (pixels) and $A,B,C$ are constants.
The fitting is restricted to data that lie above and to the right of the dashed lines.
The horizontal dashed line on figure 
\ref{HST_image} shows the approximate noise level. The vertical dashed line is placed at three times the resolution of
the HST, about $3\times\ 2$ pixels.

The model curves give an empirical estimate for the lengths, $\ell$, of the snow lines in the
different directions. At a heliocentric distance of 3.8 au, the apparent
length in the solar direction is 3500 km, and 1300 km in the two perpendicular directions. At this distance, the comet is 
moving toward both the Sun and the Earth with a projection angle of 9.6$^\circ$ and a foreshortening factor
of 5.8.  Scaling the projected radius, $\rho$, implies a snow-line length in the solar direction of 29,600 km.
The characteristic length of the exponential destruction
indicates that volatile ice grains, rather than refractory dust grains, contribute most of the total scattering cross-section in the coma.

The HST observations indicate that the sunward anti-tail from 3I/ATLAS is different from the two more
commonly observed anti-tails. Ejection of gas and dust is commonly anisotropic in the direction
of the Sun, and particularly active regions of the nucleus can produce jet-like structures directed toward the Sun.
Second, the projection of the current position of a coma against a stream of previously ejected dust
may place the stream on the sunward side of the coma due to the difference in viewing perspective
at the past time of ejection and the current time.

\begin{figure}
\includegraphics[width=3.0in,trim={0 0.0in 0.0in 0.0in},clip]{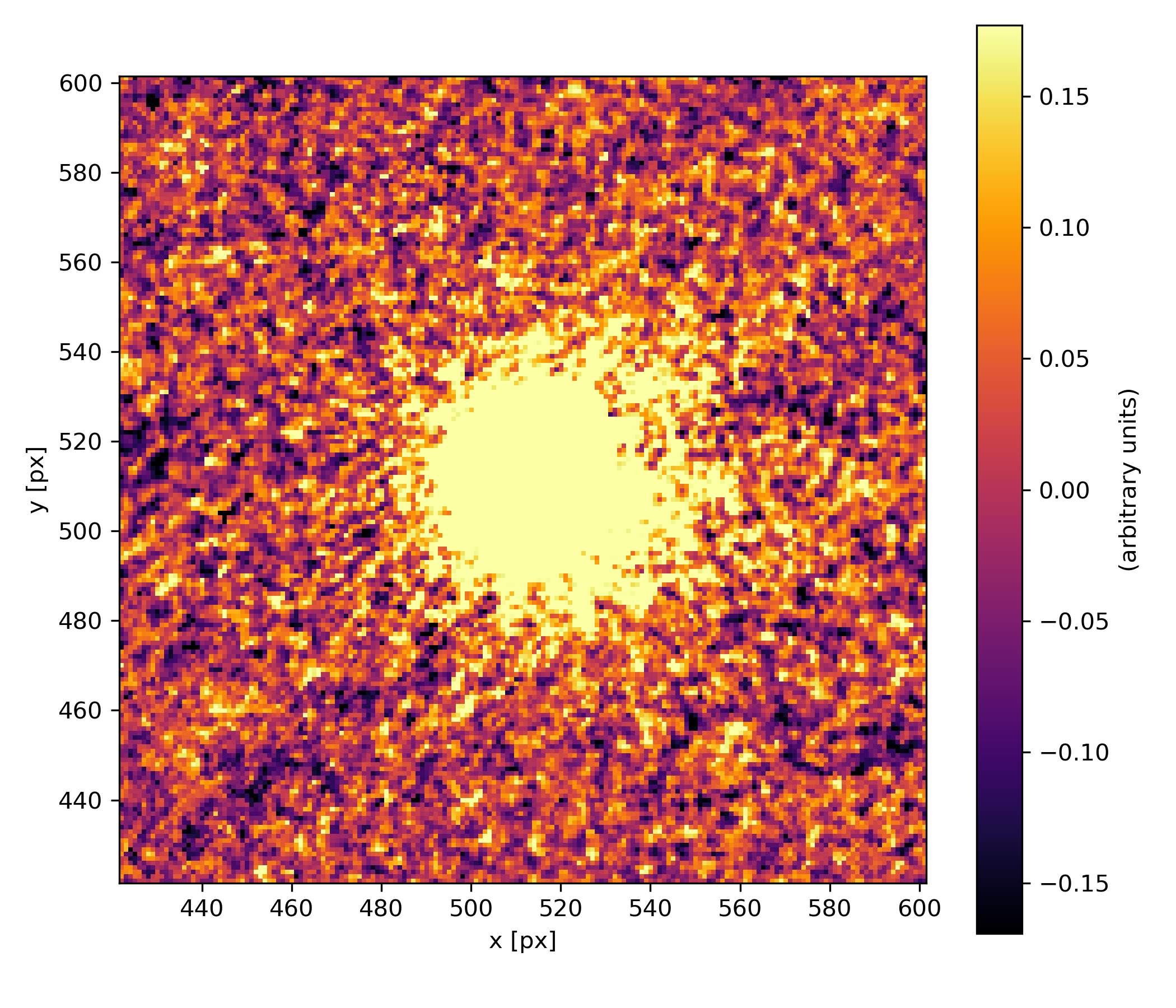} 
\caption{
HST image filtered by the rotational-gradient transform with ($\Delta r, \Delta \alpha) = (3,10)$ (equation \ref{RG} showing an
absence of small scale anisotropy. 
}
\label{rotgrad}
\end{figure}

Sunward-directed jets are characteristically narrower, shorter, and fainter than the comae in
which they are embedded. Although
they can create asymmetry in the larger comae, 
they may not be easily apparent without image enhancement, for example with the 
Larson-Sekanina rotational-gradient
method of difference imaging \citep{Larson_1984}. This filtering is scale dependent
according to the spatial separations, ($\Delta r, \Delta\alpha)$, in their equation 1,
\begin{equation}\label{RG}
\begin{split}
B^\prime(r,\alpha; \Delta r, \Delta \alpha) &= [ B(r,\alpha) - B(r-\Delta r,\alpha - \Delta \alpha)]   \\
  & \quad +[ B(x,P) - B(x-\Delta x,P + \Delta P)] 
\end{split}
\end{equation}
where $B$ is the image brightness in polar coordinates ($r,\alpha)$. Application to the HST image, 
shown in figure \ref{rotgrad},
finds no anisoptropies on the scale of $(\Delta r, \Delta \alpha) = (3,10)$ (pixels, degrees) as might
be expected for a jet-like ejection. The broader anisoptropy of the 3I/ATLAS anti-tail begins to show in the
differenced image on scales of $(\Delta x, \Delta) \alpha \sim (60,30)$.  This is consistent with the 
hypothesis that the
3I/ATLAS anti-tail is created by anisotropic destruction by sublimation of the ice grains rather than
anisotropic jet emission.

Apparent anti-tails due to projection are sometimes seen when a periodic comet crosses its aphelion orbital
point (opposition or true anomaly of 180$^\circ$). At this point the comet's orbital plane may be seen
nearly edge-on.
Large dust grains ejected at the comet's previous perihelion passage or passages, whose orbits
approximate the comet's, may appear in projection with the comet as narrow sunward tails or necklines.
In contrast, 3I/ATLAS is on the incoming leg of its hyperbolic orbit. Furthermore, the
exponential decline in the number density of scatterers indicated by the Haser-type surface brightness
profiles suggests that the scatterers
are destroyed before they have time to wrap around the comet to form a traditional antisolar tail
whose remnants might be seen in projection with the coma. 

For further comparison, the images of the comet 67P / Churyumov-Gerasimenko analyzed in \citet{Rosenbush_2026}
illustrate the three types of previously known tails, a traditional dust tail, jet-like ejection revealed by
the rotational-gradient filter, and projected necklines.
In the 6 October 2021 observations (Fig. 3), comet 67P shows a normal antisolar dust tail 
(PA $\approx 266^\circ$) and a sunward jet (PA $\approx 127^\circ$) produced by an active area on the nucleus.
In the later 6 February 2022 images (Fig. 4), a third feature appears, a bright sunward linear structure identified as a neckline, 
composed of old dust ejected near perihelion that follows the comet but on a slightly different orbital path.
At the same time, two active jets (J1, J2) are seen roughly perpendicular to the Sun-comet vector, representing current outgassing activity.
The 3I/ATLAS anti-tail does not fall easily into any one of these categories.

This discussion in this section indicates that the anti-tail from 3I/ATLAS develops from the dependence of the
length of the snow line on the illumination angle. So far, we have a mathematical fit for the radial surface brightness
in terms of the Haser model that provides estimates of the lengths but not a description of the physics. We now proceed to
estimate the snow-line length from physical principles which is the main goal of this study.

\section{A physical model for the anti-tail}

Theoretically, we expect that  the snow line or
survival length of a sublimating ice grain of size $a$, is,
\begin{equation}
\ell(a) = v(a)t_{\rm life}(a) .
\end{equation} 
This suggests a process that results in a longer survival length in the sunward direction. 
The two factors in the survival length are the velocity of the ice grains and their survival time against sublimation. 

The velocity of the ice grains is estimated with the Haser-Whipple \citep{Haser_1957,Whipple_1951} model for free molecular
acceleration,
\begin{equation}
a_d = \frac {F_{\rm drag} }{m_d} =  \frac{C_D \pi a^2  v_g J}{\rho_d a} 
\end{equation}
where
$C_D\sim 1$ is the drag coefficient; 
$v_g = \sqrt{8 k_B T /(\pi m_g)}$  ($\rm m\ s^{-1}$) is the free-expansion gas velocity where $T < 200$ K is the surface temperature ; 
and $J$ is the mass flux (kg $\rm m ^{-2} s^{-1}$) from sublimation; and $\rho_d$ is the grain density.
Integration of the acceleration with some assumptions results in the Probstein terminal velocity \citep{Finson_1968},
\begin{equation}
v_\infty \approx \mathrm{min} \bigg[ \sqrt{ \frac{3C_D}{2} } \sqrt{ \frac { v_g R_n J}   {\rho_d a}  }, v_\mathrm{g} \bigg ] .
\end{equation}
The radius of the cometary nucleus, $R_n$, is used as a characteristic length for the acceleration. This
estimate requires capping the terminal velocity at the free-expansion speed of the gas, $v_\mathrm{g}$.
The terminal velocity scales as $v_\infty \propto J^{1/2} a^{-5/4}$ up to its maximum.
This velocity gives us the first of the two factors in the survival length scale $\ell(a) = v_\infty(a) t_{\rm life}$.

The survival time of a spherical ice grain is proportional to its size,
\begin{equation}
\tau = \frac {m_g} {dm_g/dt} = \frac { \frac{4}{3}a^3\rho_d} {4\pi a^2 J} = \frac{\rho_d}{3} \frac{a}{J_d} \propto a
\end{equation}
Here $J_d$ is the sublimation mass flux off the ice grain. 
The terminal velocity and the lifetimes of the ice grains are now described as a function of the ice grain size.

Ice grain sizes are described by a power-law distribution with an index that is reliably maintained at $-3.5$ by 
fragmentation. The maximum ice grain size that can escape the gravitational pull of the nucleus is given by
Whipple's maximum liftable grain size. Although this maximum size grain
will not actually escape the nucleus since its terminal velocity is zero, we can
use this maximum
of the grain-size distribution. 
From the balance between gravity and the drag force. 
\begin{equation}
a_{\rm max} = \frac { 3C_D v_g J  } {  4 \rho_d g   }
\end{equation}
where
$g$ is the gravitational force of the nucleus, 
$R_n$ is
the nuclear radius,
$\rho_n$ is nuclear density, and 
$J$ is the mass flux off the nucleus. 
The survival length of larger grains whose terminal velocities are below
the maximum of the free-expansion velocity scales
as $a^{-1/4}$. For smaller grains that are accelerated to $v_\mathrm{g}$, the maximum velocity, $\ell \propto a$.

The ice grain sizes and hence their lifetimes and also 
their terminal velocities are both related to the mass
flux off the nucleus, $J$.

The mass flux
derives 
from the energy balance at the surface of the nucleus \citep{Prialnik_2004},
\begin{equation}\label{energy}
(1-A) \frac{ S_0} { r^2_H} \cos(\theta) = \epsilon \sigma T^4 + L_s J  - 
\Gamma \sqrt{ \omega/\pi} ~ ( T - T_b )
\end{equation}
where
$A$ is the Bond (bolometric) albedo; 
$S_0$ is the solar constant at 1 au (1361 W $\rm m^{-2}$);
$r_H$ is the heliocentric distance in au; 
$\cos(\theta)$ is the illumination factor with
$\theta$ the illumination or hour angle (HA) around the nuclear circumference and $\theta=0$ towards the Sun;
$L_s$ is the latent heat of sublimation;
$\omega$ is the angular frequency $2\pi/P$ for period $P$; and
$T_b$ is the mean equilibrium temperature. 
$T_b$ is found from the same equation by dropping the last term,
replacing $\cos \theta$ by its mean $\langle \cos \theta \rangle \approx 1/\pi$, and solving for $T$.

The left side of the equation is the power from insolation. The first term on the right
side is the radiative cooling. The second is the cooling by sublimation. The third
term is the heat flow by conduction from the nuclear interior. This maintains a minimum
temperature on the dark side of the nucleus. The form in equation \ref{energy}  is 
an approximation of $\kappa {\partial T} / {\partial z}$ for radial direction, $z$. The thermal inertia, $\Gamma$, and the
thermal conductivity, $\kappa$ are related as
$\Gamma=\sqrt{\kappa \rho_n c_p} \rm\ \  = 10 -60 \ \ (J\ m^{-2}\ K ^{-1}\ s^{-1/2})$, where $\rho_n$ is the nuclear density
\citep{Delbo_2015,Groussin_2019,Groussin_2025,Boissier_2011}.

The mass flux has to
satisfy the kinetics at the gas-solid interface described by the 
Hertz-Knudsen formula,
\begin{equation}
J(T) = \alpha \frac { m_g (P_{\rm vap}(T) - P_\infty)} { \sqrt{ m_g / (2\pi k T) } }
\end{equation}
where
$\alpha\sim 1$ is an accommodation factor.

The calculation of the mass flux, $J$, is a two-step procedure.
The temperature for the surface is obtained from the energy balance with the
HK formula for $J$.  The mass flux is then
determined for this equilibrium temperature from the HK formula.  

If the nucleus is a mixture of components, then we replace the single term, $LJ$ by
the sum of terms for the several components. We consider a comet that is 80\% H$_2$O ice
and 20\% CO$_2$ ice. Then,
\begin{equation}
LJ = L_{\rm H_2O} J_{\rm H_2O} f_{\rm H_2O} + L_{\rm CO_2}J_{\rm CO_2}f_{\rm CO_2}
\end{equation}

\begin{figure}
\includegraphics[width=3.0in,trim={0 0.0in 0.0in 0.0in},clip]{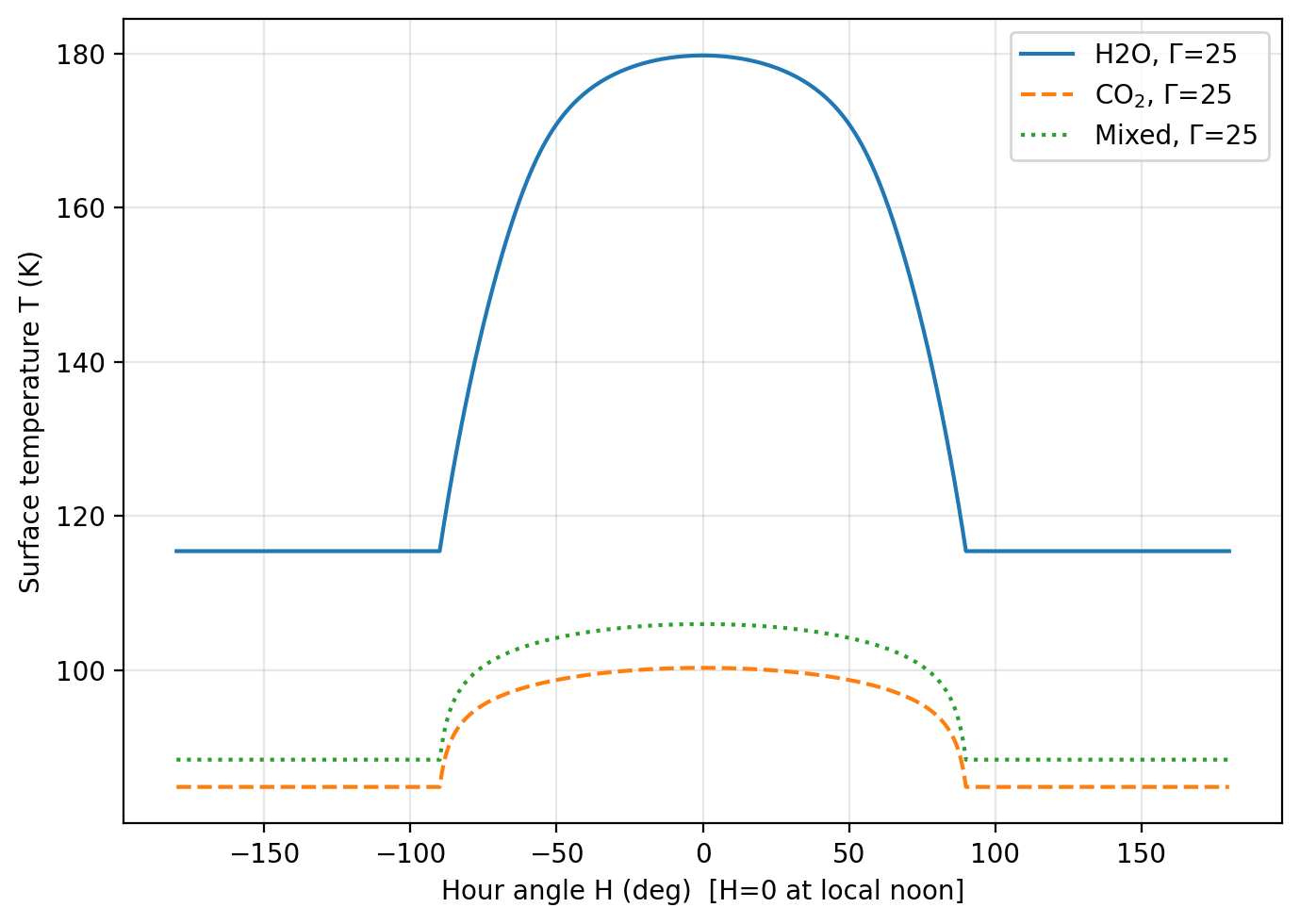} 
\caption{
Surface temperature of the nucleus as a function of angle for H$_2$O, CO$_2$, and mixed 80\% H$_2$O  20\% CO$_2$. }
\label{surface_temperature}
\end{figure}

Figure \ref{surface_temperature} shows the solutions for each of the pure species and for the mixture.
The rapid sublimation of CO$_2$ cools the surface of the mixture and keeps the temperature of the
H$_2$O too cold for sublimation as previously noted \citep{Lisse_2025}. 
The gas in the coma is then predominantly CO$_2$
as observed \citep{Lisse_2025,Cordiner_2025}.  Since the surface temperatures are too low for
a significant sublimation mass flux of H$_2$O, the relatively low proportion of gas phase H$_2$O relatve to CO$_2$
in the coma
\citep{Cordiner_2025} may derive from the sublimation
of the ice grains ejected by gas drag.

\begin{figure}
\includegraphics[width=3.0in,trim={0 0.0in 0.0in 0.0in},clip]{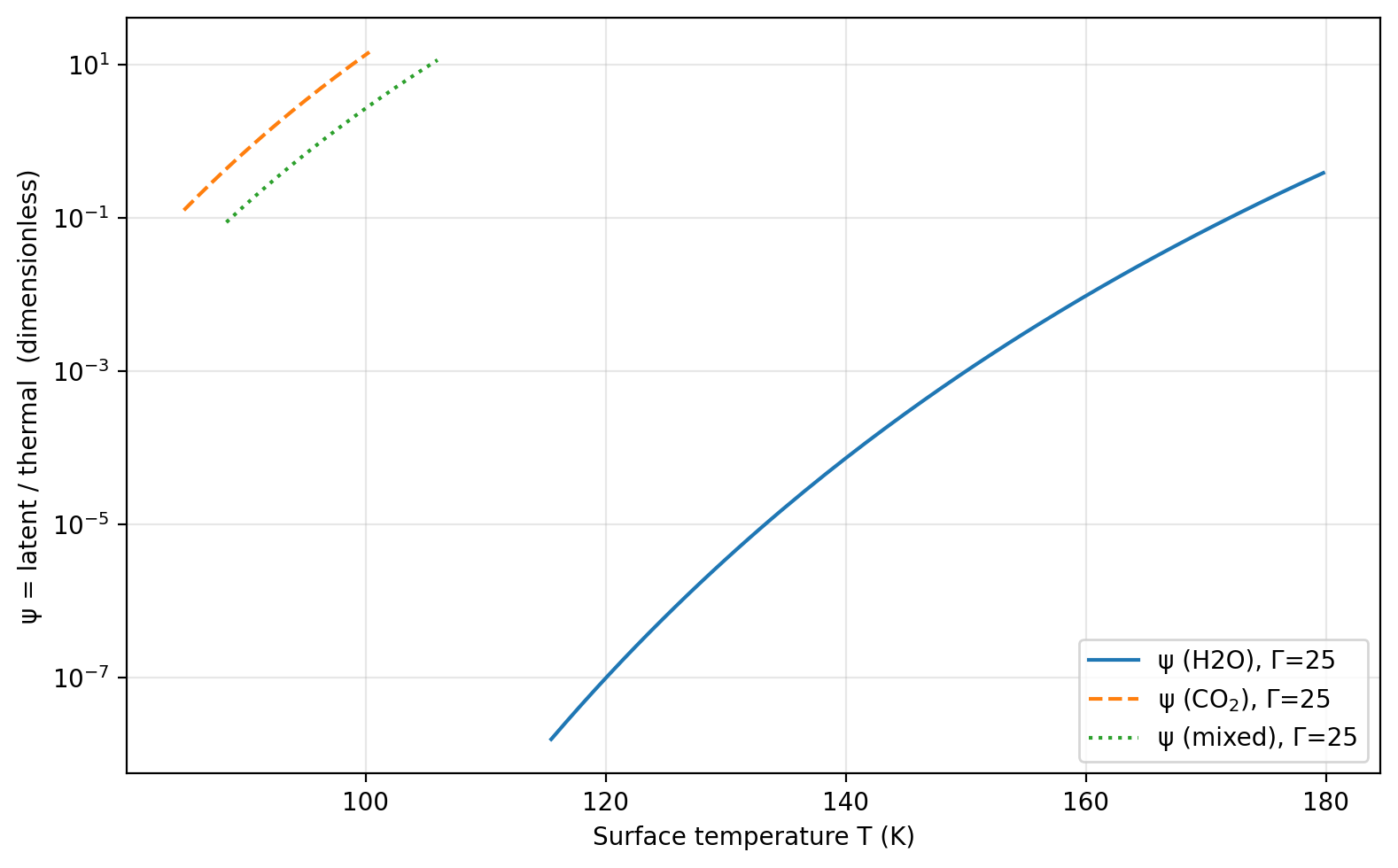} 
\caption{
 Ratio $\psi$ (equation \ref{psi}) }
\label{ratio_psi}
\end{figure}

The thermal response time scale 
may be estimated as,
\begin{equation}
\tau(T) = \frac{ \Gamma \sqrt{P/\pi} } { 4\epsilon\sigma T^4 + \Sigma f_i L_i dJ/dT }
\end{equation}
The response time varies from a minimum of 170 s at HA=0 to 3.7 hr at HA=90.0
both shorter than the 16 hour observed rotation period \citep{SantanaRos_2025,FuenteMarcos_2025}.

The thermal response time scale is longer at lower temperatures. Because of the thermal lag, the temperature
on the dark side of the comet will be warmer than its equilibrium temperature We can estimate the surface
temperatures more precisely by including the thermal lag. Written as a finite difference approximation, we
estimate the temperature of an equatorial patch of the surface as it rotates with the comet, assuming for convenience
a rotation axis perpendicular to the Sun-comet vector (maximum rotation speed),
\begin{equation}
T_{n+1} = T_b (HA_{n+1}) + [  T_n - T_\mathrm{eq}( HA_{n+1} )  ]    e^{\Delta t/\tau_{n+1}}
\end{equation}

\begin{figure}
\includegraphics[width=3.0in,trim={0 0.0in 0.0in 0.0in},clip]{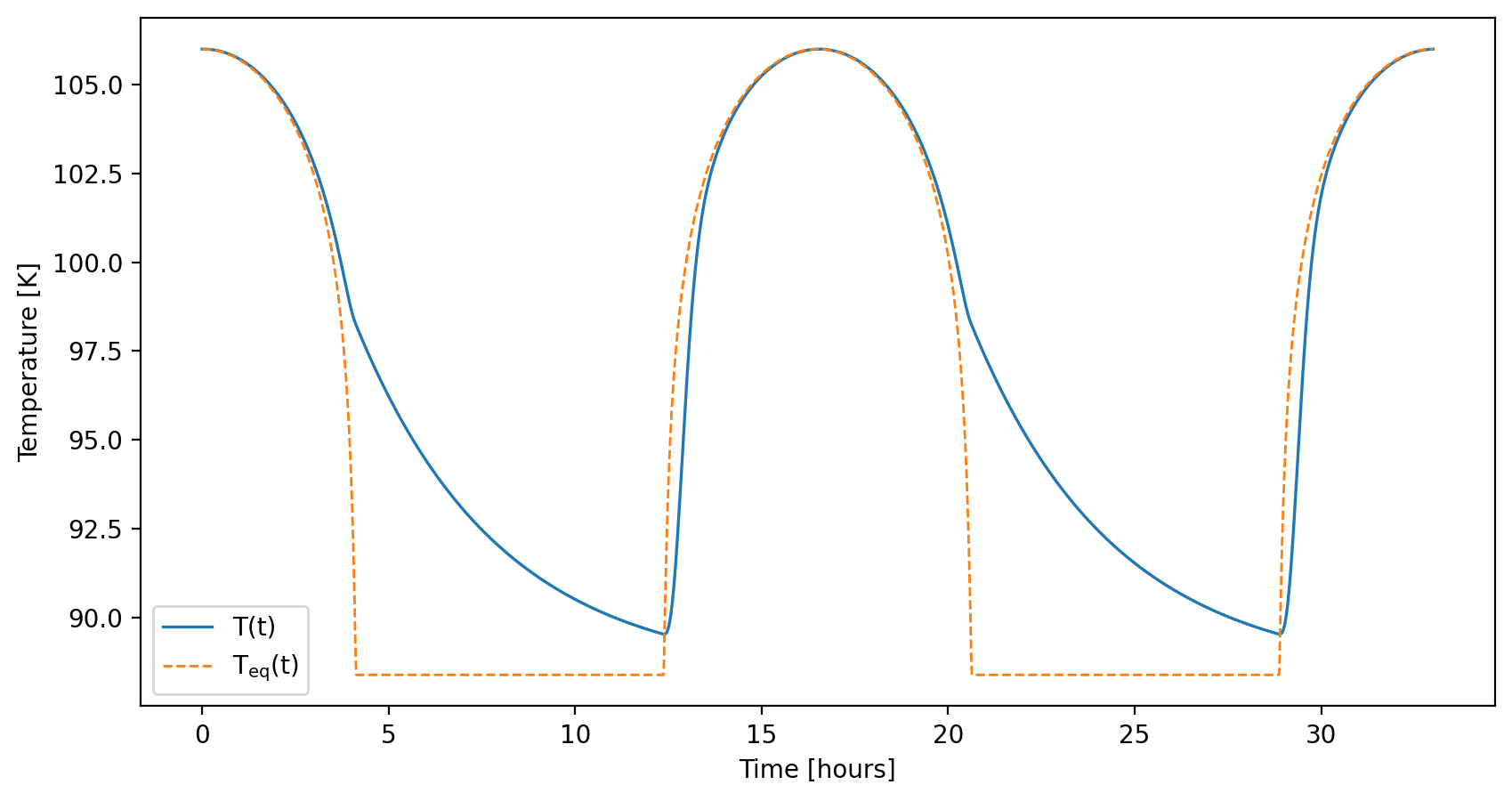} 
\caption{
Equilibrium and time-dependent surface temperatures of an equatorial patch of the mixed composition nucleus during two rotations. 
The thermal inertia, $\Gamma = 25$ J m$^{-2}$ K$^{-1}$ s$^{-1/2}$.}
\label{T_time25}
\end{figure}

\begin{figure}
\includegraphics[width=3.0in,trim={0 0.0in 0.0in 0.0in},clip]{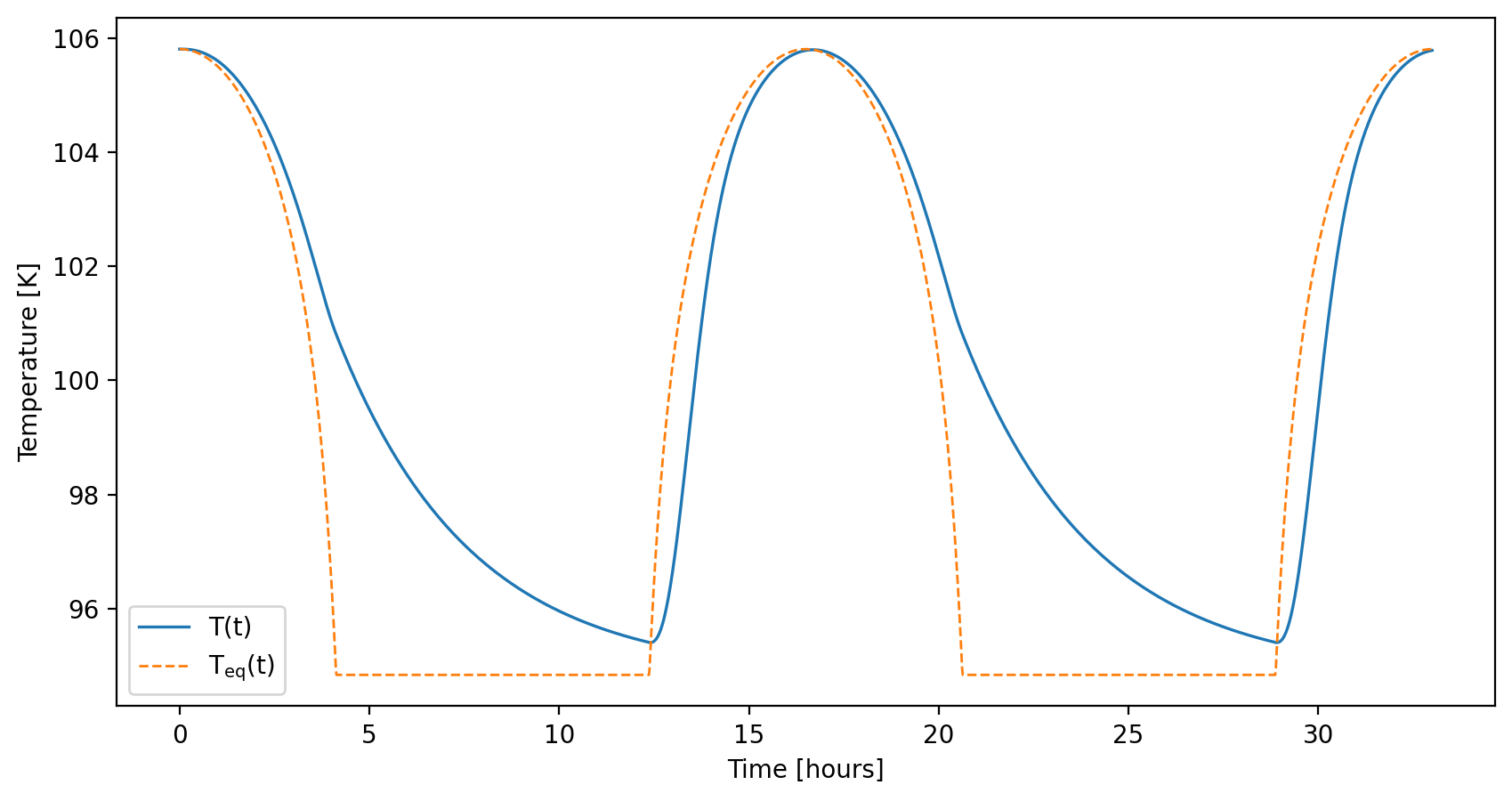} 
\caption{
Same as figure \ref{T_time25} but with the thermal inertia, $\Gamma = 100$ J m$^{-2}$ K$^{-1}$ s$^{-1/2}$.}
\label{T_time100}
\end{figure}

The time series converges immediately to a steady state and is shown in figure \ref{T_time25} for
$\Gamma=25$. For comparison, figure \ref{T_time100} shows the result for $\Gamma=100$. The 
difference amounts to an increase in the minimum temperature of ~7\% on the dark side of the comet.

The solution for the
mass flux has two limiting cases depending on whether the cooling is
dominated by radiation or sublimation.
If  $\epsilon\sigma T^4 << (1-a)S_0r^{-2} )$, 
then radiative losses are negligible and most of the solar energy goes into
sublimation. Ignoring conduction, the energy-limited mass flux is,
\begin{equation}
J_E \approx \frac {(1-a) S_0 } { Lr_H^2 } \cos\theta .
\end{equation}
In this case, the mass flux is independent of temperature.

If the sublimation is limited not by the available energy but by the kinetics at the interface, then
$J \approx J_HK$ given by the HK formula. The dependence on $T$
is more complex since the saturation
vapor pressure is given by the Clausius-Clapeyron equation,
\begin{equation}
P_{\rm vap} \approx P_0 \exp (-L/RT).
\end{equation}
In this case $J$ has a strong, exponential dependence on temperature, $J \sim T^{-1/2} \exp(-L/RT)$.

We can determine whether the nuclear surface is in the energy or kinetic-limited regime as a function of 
angle or temperature for a given composition. Define the radiative temperature,
\begin{equation}
T_{\rm rad} = (F_{\rm abs} /\epsilon\sigma)^{1/4}
\end{equation}
where $F_{\rm abs}$ is the absorbed solar flux.
For each species define the function,
\begin{equation}\label{psi}
\psi_i = \frac{  f_i L_i J_i (T_{\rm rad}) } { F_{\rm abs}} .
\end{equation}
If $\Sigma_i \psi_i >> 1$: energy limited -- sublimation cooling dominates $T < T_{\rm rad}$.  $J$ independent of surface T. \\
If $\Sigma_i \psi_i << 1$: kinetic limited -- radiation dominates $T\approx T_{\rm rad}$ and $J \sim T^{-1/2} \exp(L_i/RT)$. \\

Figure \ref{surface_temperature} (right)
shows that $\psi$ crosses the critical value for CO$_2$ and for the mixture, but remains $<1$ for H$_2$O for almost 
the full range of surface temperatures. 

\begin{figure}
\includegraphics[width=3.0in,trim={0 0.0in 0.0in 0.0in},clip]{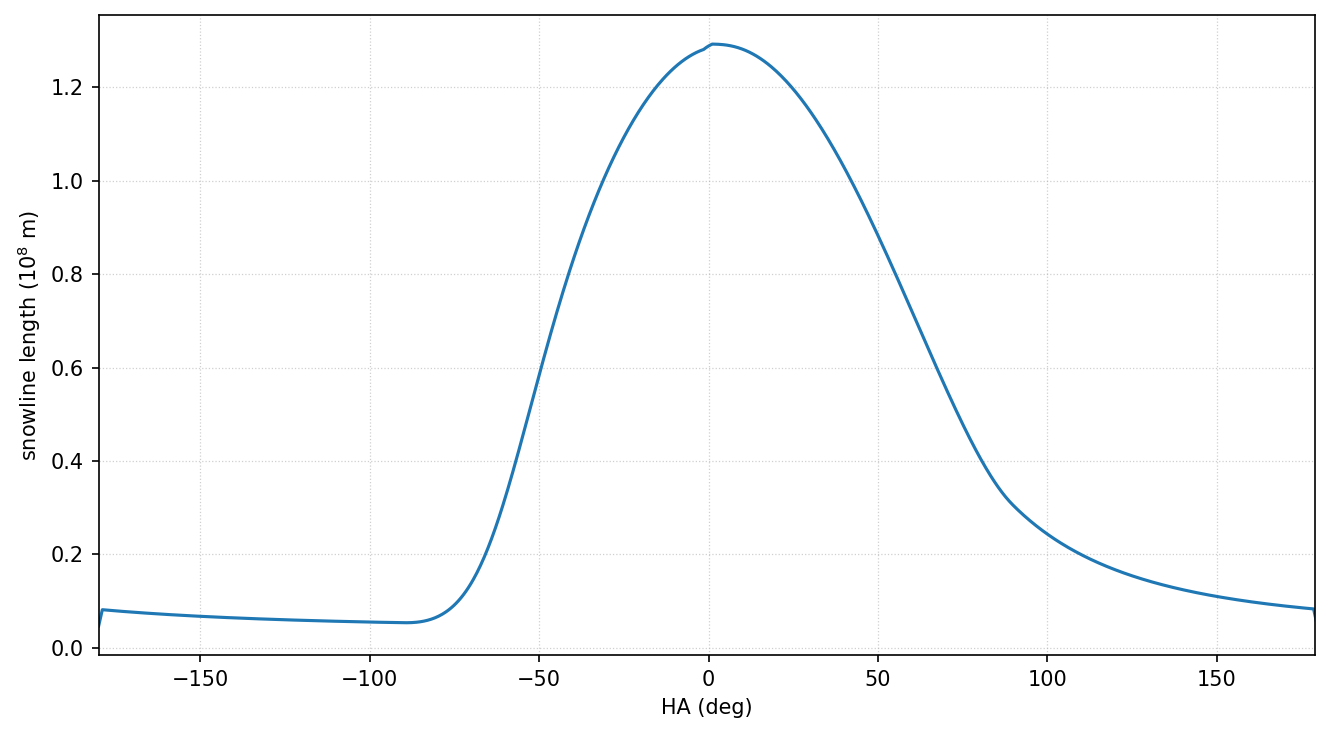} 
\caption{
Snow line for water ice as a function of nuclear hour angle for three sizes of ice grains.
}
\label{snowline}
\end{figure}

Figure \ref{snowline} shows the length of a representative snow line as a function of hour angle. The
figure shows that a representative snow line varies in length by a factor of about seven over the
range of hour angle $\pm 90^\circ$. This variation creates the sunward anti-tail.

The
maximum of the grain size distribution varies by about two magnitudes as a function of the hour angle. 
The snow line shown is generated by multiplying the maximum grain size by a constant factor of $ f=10^{-5}$
so that the grain sizes as a function of hour angle are centered around the optical wavelength, $0.5\ \mu$m.
 Ice grains of this size dominate the total scattering cross-section. 
Thus the snow line shown is $\ell(HA) = v_\infty(fa_\mathrm{max}(HA))~ t_\mathrm{life}(fa_\mathrm{max} (HA))$.

Aside from the asymmetry produced by the thermal lag, 
the curves for the snow line have the same functional dependence on hour angle as the maximum
ice grain size and the sublimation mass flux of CO$_2$. All are well-fit by a cosine function implicating
the illumination angle in the energy balance as the dominant effect on the length of the snow line.

\section{Discussion}

The model of the anti-tail as a result of the snow-line length in a coma composed of CO$_2$ gas,
H$_2$O ice grains is an alternative to other
hypotheses such as an anti-tail from enhanced ejection of large dust grains followed by fragmentation
\citep{Cordiner_2025}; from sunward jets \citep{Jewitt_2025b} and/or shaped by solar radiation pressure \citep{Jewitt_2025}; or,
from anisotropic dust emission from a nucleus with an in-plane rotation pole \citep{Chandler_2025}.
The snow-line model aligns with other observations. In particular, the surface temperature of the nucleus
maintained by the sublimation of CO$_2$ prevents significant sublimation of H$_2$O off the nucleus.
Instead, the H$_2$O gas in the coma derives from the sublimation of the ice grains.  This allows
a comet with a volatile composition of 4:1 H$_2$O:CO$_2$ to produce a coma with a 1:8 composition
in the gas phase \citep{Keto_2025b} as observed by the JWST \citep{Cordiner_2025}. The snow-line
model also allows for the disappearance 
of the anti-tail at heliocentric distances within $\sim 3.5$ au where the ice coma collapses
due to higher insolation and rapid destruction of the ice grains \citep{Keto_2025b} as observed
\citep{Hoogendam_2025,NOIRLab2525} .

\section{Conclusions}

We compare simple physical models of cometary coma with observations from the Hubble Space Telescope.
The Haser-type outflow solution with exponential destruction (equations \ref{haser1} - \ref{haser3}) 
provides good fits to the observed surface brightness profiles in the
solar and perpendicular
directions, indicating that the sublimation of ice grains in the coma, driven by the sublimation mass flux of CO$_2$ gas from the nucleus, 
determines the observed curvature of the profiles and the angular anisotropy. 

Within a radial model, the anti-tail emerges naturally as a consequence of anistropic illumination.
We find that the angular dependence of 
the characteristic length scale for the destruction is primarily due to the illumination angle of the cometary surface which affects the
sublimation mass flux of CO$_2$ gas off the cometary surface. The mass flux determines
both the terminal velocity of the ice grains and the maximum size in the grain size distribution and therefore the
characteristic survival length of the ice grains in the outflow. 

The development of an observable anti-tail depends on a favorable combination of cometary composition and 
solar insolation that both affect the
sublimation mass flux off the cometary surface as well as the destruction of the ice grains by sublimation. 
In particular,
the total scattering cross-section should be dominated by volatile ice grains. The sublimation mass flux and radiation
pressure should not be
so large as to create a traditional cometary tail, both indicating that an anti-tail is more likely to be observed at  
larger heliocentric distances. This in turn requires
high-angular-resolution observations, such as those of the HST. As 3I/ATLAS approaches the Sun, the physical conditions and shape of the coma will change.  

\section{Acknowledgements}  
AL was supported in part by the Galileo Project and the Black
Hole Initiative. The authors appreciate the help of Prof. D. Jewitt who provided the HST image.

\section*{Data Availability}

Image data provided courtesy of Prof. D. Jewitt.



\bibliographystyle{mnras}
\bibliography{comets_bib}{}



\bsp	
\label{lastpage}
\end{document}